# Variational Functionals for Excited States

Naoum C. Bacalis
*Theoretical and Physical Chemistry Institute, National Hellenic Research Foundation,
Vasileos Constantinou 48, GR-11635 Athens, Greece*

Functionals $\Omega_n$ that have local minima at the excited states of a non degenerate Hamiltonian are presented. Then, improved mutually orthogonal approximants of the ground and the first excited state are reported.

PACS: 31.15.xt, 31.15.-p, 31.10.+z, 02.70.?c

In the following the Hamiltonian expectation value of a trial wave function, $\phi$, is denoted by $E\phi$ and is called energy of $\phi$. The Hamiltonian eigenfunctions (assumed non-degenerate) are denoted by using the symbol $\psi$. All functions are assumed real and normalized.

According to the Hylleraas, Undheim, and McDonald [HUM] theorem[1] the higher roots of the secular equation tend to the excited state energies from *above*. But it should be observed that among all functions $\phi_1$, which are orthogonal to an available ground state approximant $\phi_0$, the Gram – Schmidt orthonormal to $\phi_0$

$$\phi_1^+ \equiv \frac{\psi_1 - \phi_0 \langle \psi_1 | \phi_0 \rangle}{\sqrt{1 - \langle \psi_1 | \phi_0 \rangle^2}}$$

which is the *closest*[2] to the exact $\psi_1$ (i.e. with the largest projection $\langle \psi_1 | \phi_1 \rangle^2$ - not decreased by the presence of any other components) lies energetically *below the exact* $E\psi_1$, only if $E\phi_0 < E\psi_1$:

$$E\phi_1^+ = E\psi_1 - \frac{(E\psi_1 - E\phi_0)\langle \psi_1 | \phi_0 \rangle^2}{1 - \langle \psi_1 | \phi_0 \rangle^2} < E\psi_1 ,$$

Therefore, the 2$^{nd}$ HUM root, $\phi_1^{HUM}$, lying higher than $\psi_1$, $E\phi_1^{HUM} > E\psi_1$, is necessarily *not* the closest to $\psi_1$ (while orthogonal to $\phi_0$).

On the other hand, minimizing the energy orthogonally to the available $\phi_0$, does not lead to the *closest* either: Passing through $E\phi_1^+$, it leads to an *even lower* energy: Because for any $\phi_1^{\perp+}$, chosen simultaneously orthogonal to both $\phi_0$ and $\phi_1^+$, the Hamiltonian opens the energy gap between $E\phi_1^{\perp+}$ and $E\phi_1^+$, so that, the lowest of the Hamiltonian eigenfunctions $\Psi^-, \Psi^+$, (both orthogonal to $\phi_0$) on the subspace of $\{\phi_1^{\perp+}, \phi_1^+\}$, lies lower than $E\phi_1^+$, i.e. $E\Psi^- < E\phi_1^+ < E\psi_1$, so that the lowest, $\phi_1^{MIN}$, of all such $\Psi^-$ s, obtained by minimizing the energy orthogonally to $\phi_0$, lies even lower than $E\phi_1^+$. Therefore, $\phi_1^{MIN}$ is *not* the closest to $\psi_1$ either (while orthogonal to $\phi_0$). (In fact, an appropriate sum $\Psi = \Psi^- \sqrt{\frac{E\Psi^+ - E\psi_1}{E\Psi^+ - E\Psi^-}} \pm \Psi^+ \sqrt{\frac{E\psi_1 - E\Psi^-}{E\Psi^+ - E\Psi^-}}$, orthogonal to $\phi_0$, has energy $E\Psi = E\psi_1$, with $\langle \psi_1 | \Psi \rangle^2$ not necessarily large.)

Thus, seeking $\phi_1$, approximant to $\psi_1$, orthogonal to an approximant $\phi_0$, either by the HUM theorem or by orthogonal optimization, does neither lead to $\phi_1^+$, the *closest* to $\psi_1$, nor does it raise the energy going from $\phi_1^+$ to $\psi_1$ (which is orthogonal to $\psi_0$, not to $\phi_0$). As Shull and Löwdin[3] have shown, the excited states can be calculated without knowledge of $\psi_0$. Therefore, a variational functional for $\phi_1$ would be desirable, that leads to $\psi_1$ not necessarily orthogonally to the available $\phi_0$, allowing subsequent improvement of $\phi_0$ orthogonally to $\phi_1$:

*Construction*: For a non-degenerate Hamiltonian of (unknown) bound eigenstates of a specific type of symmetry, $\psi_0, \psi_1$, and eigenenergies $E\psi_0 < E\psi_1 < ...$, a normalized approximant of $\psi_n$ can be expanded as

$$\phi_n = \sum_{i<n} \psi_i \langle \psi_i | \phi_n \rangle + \psi_n \sqrt{1 - \sum_{i<n} \langle \psi_i | \phi_n \rangle^2 - \sum_{i>n} \langle \psi_i | \phi_n \rangle^2} + \sum_{i>n} \psi_i \langle \psi_i | \phi_n \rangle \qquad (1.1)$$

where the overlap coefficients are small. The energy is

$$E\phi_n = E\psi_n - \sum_{i<n}(E\psi_n - E\psi_i)\langle\psi_i|\phi_n\rangle^2 + \sum_{i>n}(E\psi_i - E\psi_n)\langle\psi_i|\phi_n\rangle^2 \equiv E\psi_n - P_L + P_H, \quad (1.2)$$

an $n$-order saddle point, where the lower and higher than-$n$ parts, $P_L$ and $P_H$, are positive (so that $E\psi_n - P_L \le E\phi_n \le E\psi_n + P_H$ ).

The minimum of the following paraboloid, defined by

$$E\psi_n + P_L + P_H = E\phi_n + 2P_L \quad (1.3)$$

determines $\phi_n \to \psi_n$, in terms of the lower than-$n$ information ($P_L$). An expression for the behaviour of $P_L$ can be found by first considering, to leading order in coefficients, the overlap and the Hamiltonian matrix elements in terms of the (similarly predetermined as described here) approximants $\phi_i, i < n$ :

$$\langle\phi_i|\phi_n\rangle = \langle\psi_i|\phi_n\rangle + \langle\psi_n|\phi_i\rangle + \cdots$$
$$\langle\phi_i|H|\phi_n\rangle = E\psi_i\langle\psi_i|\phi_n\rangle + E\psi_n\langle\psi_n|\phi_i\rangle + \cdots . \quad (1.4)$$

Substituting $\langle\psi_i|\phi_n\rangle$ from Eqs. (1.4) to each term of $P_L$ in Eq. (1.2) gives, to leading order,

$(E\psi_n\langle\phi_i|\phi_n\rangle - \langle\phi_i|H|\phi_n\rangle)^2/(E\psi_n - E\psi_i)$, which suggests an examination, in terms of known quantities, of the expression

$\sum_{i<n}\left[(E\phi_n\langle\phi_i|\phi_n\rangle - \langle\phi_i|H|\phi_n\rangle)^2/(E\phi_n - E\phi_i)\right]$ . This, as directly verified, when both $\phi_i = \psi_i$ and [in Eq.(1.2)] $P_H \to 0$ ,

reduces to $P_L\left(1 - \sum_{i<n}\langle\phi_i|\phi_n\rangle^2\right)$ . Therefore, for $P_H \ne 0$ the behaviour of the paraboloid of Eq. (1.3) close to $\psi_n$ is reasonably

described by the functional $\Omega_n$ :

$$E\psi_n + P_L + P_H = E\phi_n + 2P_L \to \Omega_n \equiv E\phi_n + 2\frac{\displaystyle\sum_{i<n}\frac{(E\phi_n\langle\phi_i|\phi_n\rangle - \langle\phi_i|H|\phi_n\rangle)^2}{E\phi_n - E\phi_i}}{1 - \displaystyle\sum_{i<n}\langle\phi_i|\phi_n\rangle^2} \quad (1.5)$$

with a local minimum at $\phi_n = \psi_n$ , which is paraboloidal, by construction, when $\phi_i = \psi_i$ .

*Proof*: $\Omega_n$ has a true local minimum at $\phi_n = \psi_n$ when $\phi_i$ are *approximants* of $\psi_i$ ($\phi_i \approx \psi_i$), while $E\phi_n$ has a saddle point there: By collecting the contribution of the higher than-$n$ subspace for each $\phi_i$ wave function , $i \le n$ , to the contribution of a normalized function $\phi_i^{\perp\{n\}}$, $i \le n$ , orthogonal to all lower than-$n$ $\psi_i$ eigenfunctions, i.e.

$$\phi_i^{\perp\{n\}} = \sum_{j>n}\psi_j\langle\psi_j|\phi_i\rangle \Big/ \sqrt{\sum_{j>n}\langle\psi_j|\phi_i\rangle^2} , \quad i \le n , \quad (1.6)$$

where the overlap and Hamiltonian matrix elements are generally non-zero, $\langle\phi_i^{\perp\{n\}}|\phi_j^{\perp\{n\}}\rangle \ne 0$ , $\langle\phi_i^{\perp\{n\}}|H|\phi_j^{\perp\{n\}}\rangle \ne 0$ , $i, j \le n$ , and whose energies, obviously, are $E\phi_i^{\perp\{n\}} > E\psi_n$, $i \le n$ , it is directly verified that all the principal minors $A_n^i$ , $i \le n$ , of the Hessian determinant $A_n^n$ of $\Omega_n$ , along the main diagonal, i.e. those which are required by the second derivatives theorems of calculus (Sylvester's theorem), are, at the desired place $\phi_n = \psi_n$ , $\phi_i \ne \psi_i$ , $i < n$ , positive, if $\phi_i$ are close to $\psi_i$: Each principal minor determinant (denoted by the main diagonal)

$$A_n^{k<n} \equiv Det\left[\frac{\partial^2\Omega_n}{\partial\langle\psi_0|\phi_n\rangle\partial\langle\psi_0|\phi_n\rangle} \cdots \frac{\partial^2\Omega_n}{\partial\langle\psi_i|\phi_n\rangle\partial\langle\psi_i|\phi_n\rangle} \cdots \frac{\partial^2\Omega_n}{\partial\langle\psi_k|\phi_n\rangle\partial\langle\psi_k|\phi_n\rangle}\right]_{\phi_n=\psi_n,\,\phi_i\ne\psi_i,\,i<k}$$

equals

$$A_n^{k<n} = 2^k\prod_{i=0}^{k}(E\psi_n - E\psi_i) > 0 \ (+O[\langle\psi_q|\phi_r\rangle\langle\psi_s|\phi_t\rangle]) . \quad (1.7)$$

where there are no coefficients (which depend on the quality of $\phi_i$ ) of 1$^{st}$ power, while the Hessian itself

$$A_n^n \equiv Det\left[\frac{\partial^2 \Omega_n}{\partial\langle\psi_0|\phi_n\rangle\partial\langle\psi_0|\phi_n\rangle}\cdots\frac{\partial^2 \Omega_n}{\partial\langle\psi_i|\phi_n\rangle\partial\langle\psi_i|\phi_n\rangle}\cdots\frac{\partial^2 \Omega_n}{\partial\langle\phi_n^{\perp\{n\}}|\phi_n\rangle\partial\langle\phi_n^{\perp\{n\}}|\phi_n\rangle}\right]_{\phi_n=\psi_n,\,\phi_i\neq\psi_i,\,i<n}$$

equals

$$A_n^n = 2^n \left(E\phi_n^{\perp\{n\}} - E\psi_n\right)\prod_{i=0}^{n-1}\left(E\psi_n - E\psi_i\right) > 0 \; \left(+O\left[\langle\psi_q|\phi_r\rangle\langle\psi_s|\phi_t\rangle\right]\right). \tag{1.8}$$

If $\phi_i$ are close to $\psi_i$, all these determinants of Eqs. (1.7 - 1.8) are positive, hence the Hessian matrix is positive definite, therefore, the functional $\Omega_n$ has a local minimum at $\phi_n = \psi_n$, which determines $\psi_n$ if all $\phi_i$ approximants of $\psi_i$, $i \leq n$, are known. Obviously, $\Omega_0$ reduces to the Eckart[4] theorem for $\psi_0$.

The functional $\Omega_n$ passes from all $\psi_i$. A way to identify the desired $\psi_n$ for atoms and for diatomic molecules, is to expand (for atoms) in a basis of Slater type exponentials whose prefactors are *not monomials*, but rather they are variationally optimized *polynomials*: initially starting from the *identifiable* associated Laguerre polynomials, because these are *not severely modified* during optimization; Also identifiable (for diatomic molecules) are the (separable into radial and angular parts) variationally optimized *one*-electron-*diatomic-molecule*-type orbitals. Both significantly reduce the size of a configuration interaction expansion. [5]

*Improving $\phi_0$ orthogonally to $\psi_1$*: If $\psi_1$ were known it would be possible to improve $\phi_0$ orthogonally to $\psi_1$:[2] On the subspace of $\{\phi_0, \psi_1\}$ the highest Hamiltonian eigenvector, $\Psi^+$, is

$$\Psi^+ = \psi_1.$$

The lowest, $\Psi^-$, is orthogonal to $\psi_1$,

$$\Psi^- = \phi_0^+ \equiv \frac{\phi_0 - \psi_1\langle\psi_1|\phi_0\rangle}{\sqrt{1-\langle\psi_1|\phi_0\rangle^2}}$$

with energy

$$E\phi_0^+ = E\phi_0 - \frac{(E\psi_1 - E\phi_0)\langle\psi_1|\phi_0\rangle^2}{1-\langle\psi_1|\phi_0\rangle^2} \leq E\phi_0 \tag{1.9}$$

(same or better than $\phi_0$). Further, rotating $\phi_0^+$ around $\psi_1$ improves $\phi_0^+$ as follows: After introducing (e.g. by one more configuration) a function $\phi_0^{(2+)}$ orthogonal to both $\{\phi_0^+, \psi_1\}$, then, in the subspace of $\{\phi_0^+, \phi_0^{(2+)}\}$ (both orthogonal to $\psi_1$), the lowest Hamiltonian eigenvector $\Psi^- \equiv \phi_0^-$ has energy $E\phi_0^- \leq E\phi_0^+$, closer to $E\psi_0$, because the Hamiltonian opens the energy gap between $\{E\phi_0^+, E\phi_0^{(2+)}\}$ (in a 3-dimensional function space $\{\psi_0, \psi_1, \psi_k\}$ this would be exactly $E\psi_0$ as it can be directly verified). $E\phi_0^-$ can be further improved by further rotating around $\psi_1$ similarly, i.e. after introducing another function $\phi_0^{(3+)}$ orthogonal to both $\{\phi_0^-, \psi_1\}$ by calculating in the subspace of $\{\phi_0^-, \phi_0^{(3+)}\}$ (both orthogonal to $\psi_1$) the lowest eigenvector $\Psi^- \equiv \phi_0^{(2-)}$ which has energy $E\phi_0^{(2-)} \leq E\phi_0^-$ (even closer to $E\psi_0$); and so on.

*Improving $\phi_0$ orthogonally to $\phi_1$*: Since $\psi_1$ is never exactly known, then, it may still be possible to improve $\phi_0$ orthogonally to $\phi_1$, the best available approximant of $\psi_1$, by first computing $\phi_0^+$ orthogonal to $\phi_1$,

$$\phi_0^+ \equiv \frac{\phi_0 - \phi_1\langle\phi_1|\phi_0\rangle}{\sqrt{1-\langle\phi_1|\phi_0\rangle^2}} \tag{1.10}$$

if the condition

$$E\phi_0^+ = \frac{E\phi_0 + E\phi_1\langle\phi_1|\phi_0\rangle^2 - 2\langle\phi_0|H|\phi_1\rangle\langle\phi_1|\phi_0\rangle}{1-\langle\phi_1|\phi_0\rangle^2} \leq E\phi_0 \tag{1.11}$$

is attainable. Indeed, by expanding about $\psi_1$, as directly verified, this condition, to leading order, reads

$(E\psi_1 - E\psi_0)(1-\langle\psi_1|\phi_0\rangle^2) \geq (E\phi_0^{\perp\{1\}} - E\psi_0)\langle\phi_0^{\perp\{1\}}|\phi_0\rangle^2$, which is not impossible. Here [c.f. Eq. (1.6)] $\phi_0^{\perp\{1\}}$ is the normalized

function, orthogonal to both $\{\psi_0, \psi_1\}$, collecting all higher than-1 terms of $\phi_1$. For $\phi_0, \phi_1$ very close to $\psi_0, \psi_1$, as directly verified by expanding about $\psi_0, \psi_1$ the condition is satisfied when $\langle \psi_0 | \phi_1 \rangle^2 \leq \langle \psi_1 | \phi_0 \rangle^2$ (indicative of the relative quality of the approximants). Incidentally, all other (small) components (out of the plane of $\psi_0, \psi_1$) are less relevant, so that the opposite procedure of optimizing $\phi_1$ orthogonally to $\phi_0$ can lead to $\phi_1^{MIN}$ unpredictably *far* from $\psi_1$ with *still* $E\phi_1^{MIN} \lesssim E\psi_1$, as shown in the following example.

*Example*: Even in the subspace $\{\psi_0, \psi_1, \psi_2\}$, the orthonormal trial functions $\phi_0 = a\psi_0 + b\psi_2$, $\phi_1 = b\psi_0 - a\psi_2$ with $a = \sqrt{[(E\psi_1 - \varepsilon) - E\psi_0]/(E\psi_2 - E\psi_0)}$, $b = \sqrt{[E\psi_2 - (E\psi_1 - \varepsilon)]/(E\psi_2 - E\psi_0)}$, (small $\varepsilon$), have energies $E\phi_0 = E\psi_0 + E\psi_2 - (E\psi_1 - \varepsilon) \cong E\psi_0 + \varepsilon$ (if $E\psi_2 - E\psi_1$ is small), $E\phi_1 = E\psi_1 - \varepsilon$, while $\phi_0$ reasonably, but not particularly accurately, approximates $\psi_0$ (for instance, for He $^1$S, in a.u., $E\psi_0 = -2.903$, $E\psi_1 = -2.146$, $E\psi_2 = -2.06$, $\phi_0 = 0.9476\ \psi_0 + 0.3194\ \psi_2$ has $E\phi_0 = -2.817$ and $\phi_1 = 0.3194\ \psi_0 - 0.9476\ \psi_2$ has $E\phi_1 = -2.146 = E\psi_1$, while $\phi_1$ is orthogonal to both $\phi_0$ and $\psi_1$), so that, *any function orthogonal to the same* $\phi_0$ could be a minimization "result", $\phi_1^{MIN}$, with *arbitrary* $\langle \psi_1 | \phi_1^{MIN} \rangle$ and with $E\psi_1 - \varepsilon \leq E\phi_1^{MIN} \leq E\psi_1$.

*Demonstration of* $\Omega_1$: Minimization of $\Omega_1$, for the same $\phi_0$ of He, as above, by varying $\phi_1 = c\ \psi_0 + d\ \psi_2 + \psi_1\sqrt{1 - c^2 - d^2}$, yields $c < tol = 10^{-8}$, $d < tol$, with $E\phi_1 = -2.146$ [so that $\phi_1 = \psi_1$ and, from Eq. (1.10), $\phi_0^+ = \phi_0$].

*Further improvement of* $\phi_0$: If $E\phi_0^+ \leq E\phi_0$ [Eq. (1.11)], then, by rotating around $\phi_1$, as described above [after Eq. (1.9)], since the Hamiltonian always opens the energy gap between mutually orthogonal functions (all orthogonal to $\phi_1$), $\phi_0^+$ can be further improved (until $\langle \psi_0 | \phi_1 \rangle^2 > \langle \psi_1 | \phi_0 \rangle^2$), by always taking the lowest current eigenfunction $\phi_0^{(m-)} = \Psi^-$. At any step, $\phi_0^{(m-)}$ can be used as a new $\phi_0$ to improve $\phi_1$ via $\Omega_1$ of Eq. (1.5). In the above example of He, rotating $\phi_0$ around $\phi_1$, gives $\phi_0^{(1-)} = \Psi^- = \psi_0$ (and $\Psi^+ = \psi_2$).

*Technicalities*: If the higher eigenvalues approach each other, then the second derivatives diminish and the paraboloid $\Omega_n$ flattens within the tolerance criterion $\varepsilon_n$, used in the $\Omega_n$ minimization. Then it might be desirable to steepen it near the minimum. The simplest way would be to multiply $\Omega_n$ by a large number N, so as to distinguish the differences within the *same* $\varepsilon_n$. Also, it might be possible, by introducing one more variable, $E_F$, to minimize the functional $F[\Omega_n, E_F] \equiv \Omega_n + \left| \frac{\Omega_n - E_F}{E_F T} \right|$ : if $T$ is chosen in the order of $\Omega_n$'s curvature radius at $\psi_n$, ~(inverse of second derivatives, estimated by the Hessian minors, or by trial), then, as directly verified by expanding about $\psi_n$, $F$ is a paraboloid with minimum at $\phi_n = \psi_n$ with $F[E\psi_n, E\psi_n] = E\psi_n$.

*Summary*: $\Omega_n$ [Eq. (1.5)] has a local minimum at the excited state $\psi_n$, where $\Omega_n = E\psi_n$ and $\phi_n = \psi_n$. If $\phi_1$ is a better approximant to $\psi_1$ than $\phi_0$ is to $\psi_0$ [i.e. if, from Eq. (1.11), $E\phi_0^+ \leq E\phi_0$], then $\phi_0$ can be improved orthogonally to $\phi_1$.

---